\documentclass{article}

\PassOptionsToPackage{numbers, sort&compress}{natbib}

    \usepackage[preprint]{neurips_2025}

\usepackage[utf8]{inputenc} 
\usepackage[T1]{fontenc}    
\usepackage{hyperref} 
\usepackage{url}            
\usepackage{xurl}           
\usepackage{booktabs}       
\usepackage{amsfonts}       
\usepackage{nicefrac}       
\usepackage{microtype}      
\usepackage{xcolor}         
\usepackage{algorithm}
\usepackage{algpseudocode}
\usepackage{amsmath}
\usepackage{multicol}
\usepackage{geometry}
\usepackage{xcolor}
\usepackage{tcolorbox}
\usepackage{csquotes}
\usepackage{siunitx}
\usepackage{graphicx}
\usepackage{tikz}
\usepackage{booktabs}  
\usepackage{multirow}  
\usepackage{lineno}
\usepackage{subcaption}
\usepackage{epigraph}
\usepackage{xspace}

\definecolor{darkblue}{rgb}{0, 0, 0.5}

\newcommand{\systemname}{PrfaaS\xspace}
\newcommand{\systemnamelowercase}{prfaas}

\title{Prefill-as-a-Service: KVCache of Next-Generation Models Could Go Cross-Datacenter}

\author{
  Ruoyu Qin$^{1,2}$\quad
  Weiran He$^{1}$\quad
  Yaoyu Wang$^{1}$\quad
  Zheming Li$^{1}$\quad
  Xinran Xu$^{1}$ \\
  \bfseries Yongwei Wu$^{2}$ \quad
  Weimin Zheng$^{2}$\quad
  Mingxing Zhang$^{2}$\thanks{Corresponding author: \texttt{zhang\_mingxing@mail.tsinghua.edu.cn}} \\[6pt]
  $^{1}$Moonshot AI\qquad
  $^{2}$Tsinghua University
}

\begin{document}

\maketitle

\begin{abstract}
Prefill-decode (PD) disaggregation has become the standard architecture for large-scale LLM serving, but in practice its deployment boundary is still determined by KVCache transfer. In conventional dense-attention models, prefill generates huge KVCache traffics that keep prefill and decode tightly coupled within a single high-bandwidth network domain, limiting heterogeneous deployment and resource elasticity. Recent hybrid-attention architectures substantially reduce KVCache size, making cross-cluster KVCache transport increasingly plausible. However, smaller KVCache alone does not make heterogeneous cross-datacenter PD serving practical: real workloads remain bursty, request lengths are highly skewed, prefix caches are unevenly distributed, and inter-cluster bandwidth fluctuates. A naive design that fully externalizes prefill can therefore still suffer from congestion, unstable queueing, and poor utilization.

We present Prefill-as-a-Service (\systemname), a cross-datacenter serving architecture that selectively offloads long-context prefill to standalone, compute-dense prefill clusters and transfers the resulting KVCache over commodity Ethernet to local PD clusters for decode. Rather than treating reduced KVCache as sufficient, \systemname combines model-side KV efficiency with system-side selective offloading, bandwidth-aware scheduling, and cache-aware request placement. This design removes the requirement that heterogeneous accelerators share the same low-latency RDMA fabric, enabling independent scaling of prefill and decode capacity across loosely coupled clusters. In a case study using an internal 1T-parameter hybrid model, a \systemname-augmented heterogeneous deployment achieves 54\% higher serving throughput and 64\% lower P90 TTFT than a homogeneous PD baseline, with approximately 15\% throughput gain at equal cost, while consuming only modest cross-datacenter bandwidth.
\end{abstract}

\section{Introduction}
\label{sec:introduction}

Prefill-decode (PD) disaggregation has become the dominant deployment paradigm for large-scale LLM serving because it separates two fundamentally different phases of inference: prefill is primarily compute-intensive, whereas decode is primarily memory-bandwidth-intensive. At Moonshot AI, Mooncake~\cite{qin2024mooncake} helped push this shift into practice by treating KVCache as a first-class systems resource, a direction that has since propagated across the broader serving ecosystem via our collaboration with open frameworks such as vLLM~\cite{vllm2025}, SGLang~\cite{sglang2025}, and Dynamo~\cite{dynamo2025}. In principle, PD disaggregation should also unlock a more ambitious goal: heterogeneous serving, in which prefill runs on compute-dense accelerators and decode runs on bandwidth-optimized accelerators. Hardware roadmaps are already moving in this direction. NVIDIA's Rubin CPX~\cite{nvidia2025rubincpx} explicitly targets high-throughput long-context prefill, while architectures such as Groq's Language Processing Unit (LPU)~\cite{abts2020think, groq2025lpu} emphasize the extreme memory bandwidth required for decode. 

In practice, however, this heterogeneous vision remains difficult to realize because current PD disaggregation still relies on a strong networking assumption: when prefill and decode are placed on different nodes, the KVCache produced by prefill must be transferred quickly enough to avoid stalling computation. In conventional deployments, this effectively confines both phases to the same tightly coupled, high-bandwidth network domain, typically a single datacenter-scale RDMA fabric, where PD disaggregation works well inside a homogeneous cluster. The difficulty is that this single-datacenter paradigm does not extend naturally to heterogeneous serving. Accelerator resources are usually pooled by both chip type and physical location, so compute-oriented and inference-oriented hardware are often unavailable within the same tightly coupled domain. This creates a strong incentive to separate prefill from decode across cluster boundaries, which can reduce costs and latency for long-context requests by moving prefill to faster compute hardware. However, this advantage materializes only if KVCache transfer remains sufficiently cheap. Once prefill and decode no longer share the same high-bandwidth fabric, the generated KVCache must traverse a slower inter-cluster link. If that transfer cost is too high, it erases the prefill-side gain and becomes the new bottleneck. Even when two clusters are geographically close, requiring them to share a single RDMA-scale fabric is operationally rigid and often unrealistic. Worse, once heterogeneous deployment is forced into one tightly coupled cluster, its prefill-to-decode hardware ratio becomes difficult to adapt as traffic patterns evolve. As a result, current PD deployments still fall short of the flexibility that heterogeneous disaggregation is supposed to provide.

The central obstacle, therefore, is KVCache transfer. Recent hybrid-attention architectures change this picture in an important way. Emerging models~\cite{team2025kimilinear,qwen2026qwen3p5,xiao2026mimo,inclusion2026ring25,huang2026step,agarwal2025gpt,nvidia_nemotron_nano_v3_2025} interleave a small number of full-attention layers with a larger number of linear-complexity or bounded-state layers, such as Kimi Delta Attention (KDA)~\cite{team2025kimilinear}, Sliding Window Attention (SWA)~\cite{beltagy2020longformer}, and related mechanisms. This design substantially reduces KVCache growth relative to dense-attention architectures, often by an order of magnitude, thereby making cross-datacenter KVCache transfer plausible. But plausibility is not yet practicality: a naive design that externalizes all prefills would still suffer from bursty arrivals, skewed request lengths, uneven prefix-cache distribution, and fluctuating inter-cluster bandwidth. Hybrid architectures relax the KVCache bottleneck, but they do not eliminate the need for system design; rather, they create the opportunity that system design must exploit.

\begin{figure}[t]
    \centering
    \includegraphics[width=\linewidth]{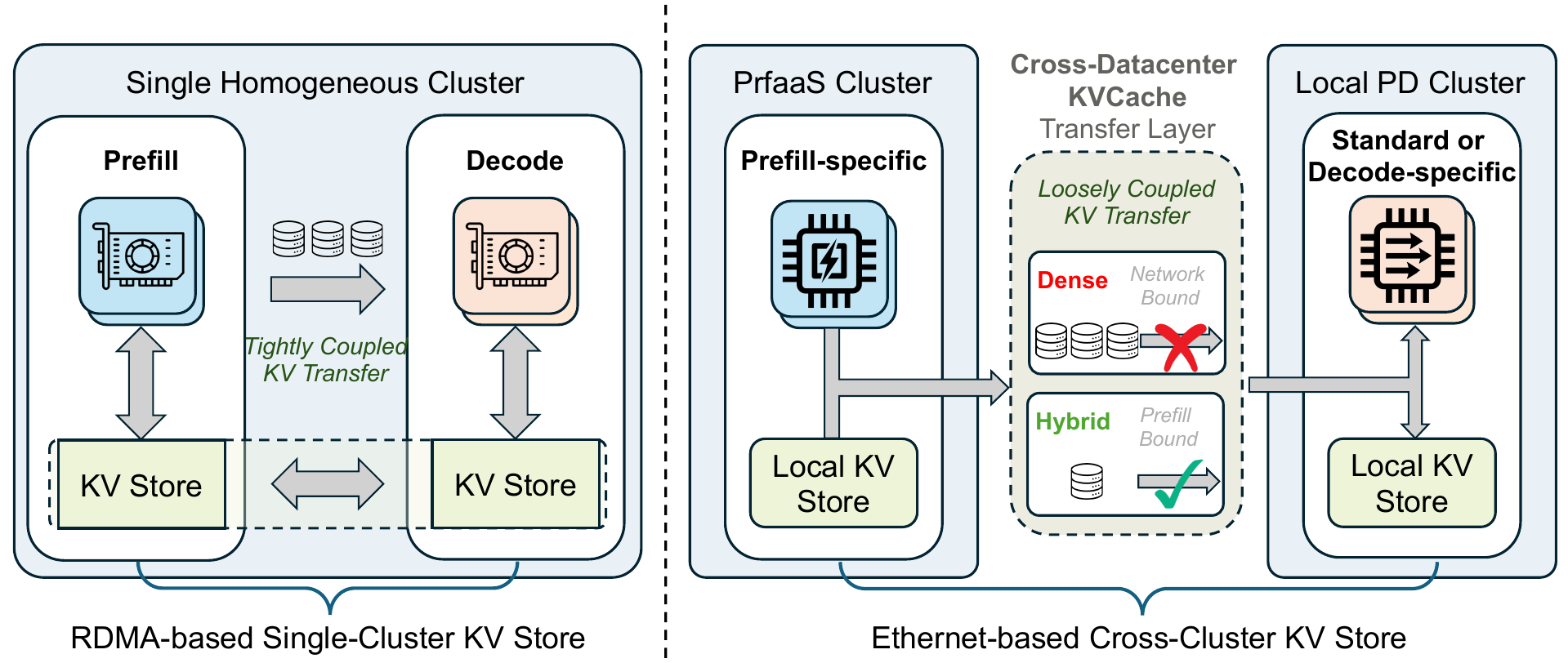}
    \begin{minipage}[t]{0.40\linewidth}
        \centering
        \subcaption{Status quo: Tightly coupled single-cluster inference.}
    \end{minipage}%
    \hfill
    \begin{minipage}[t]{0.53\linewidth}
        \centering
        \subcaption{\systemname: Multi-cluster disaggregated inference via cross-datacenter KVCache.}
    \end{minipage}
    \caption{Comparison of two deployment paradigms for PD-disaggregated LLM serving.}
    \label{fig:main_figure}
\end{figure}

This observation leads to the key design principle of this paper: \textbf{Prefill-as-a-Service (\systemname)} via cross-datacenter KVCache. As illustrated in Figure~\ref{fig:main_figure}, instead of forcing heterogeneous accelerators into a single RDMA island, \systemname constructs standalone clusters dedicated to long-context prefill using inexpensive, high-throughput compute. Besides, rather than fully separating every request, \systemname offloads only long uncached prefills to these compute-dense prefill clusters, while short requests remain on the local PD path. The resulting KVCache is then transferred over commodity Ethernet to decode-capable PD clusters. This design reflects the underlying systems reality: the motivation to split prefill is strong, but the reduced KVCache of hybrid models is still not cheap enough to justify indiscriminate transfer. What makes the design feasible is selective offloading, which concentrates cross-cluster transfer on the requests for which prefill acceleration matters most, while avoiding the inefficiency of sending short requests through a bandwidth-constrained path.

Making this design work requires scheduling and cache management that explicitly address the remaining systems challenges. Considering that bandwidth remains constrained even after KVCache reduction, \systemname uses length-based threshold routing to offload only sufficiently long requests, a bandwidth-aware scheduler to react to fluctuating link conditions before congestion accumulates, and a global KVCache manager with a hybrid prefix-cache pool to account jointly for request length, cache placement, and available cross-cluster bandwidth. These mechanisms make cross-cluster heterogeneous serving viable in realistic environments: they preserve the benefits of PD disaggregation without requiring heterogeneous accelerators to share the same low-latency RDMA fabric, and they allow compute-oriented prefill capacity and bandwidth-oriented decode capacity to scale independently across loosely coupled clusters, datacenters, or regions.

This flexibility directly addresses deployment constraints that are otherwise difficult to resolve in practice, including non-colocated accelerator classes, hardware asymmetry across cloud regions, and opportunistic remote capacity. We evaluate this idea through a case study using an internal 1T-parameter hybrid model following the Kimi Linear architecture~\cite{team2025kimilinear}. With a heterogeneous deployment consisting of a standalone \systemname cluster for long-context prefill and a conventional PD cluster for decode and short prefills, the system achieves 54\% and 32\% higher serving throughput than homogeneous PD and naive heterogeneous baselines, respectively, while consuming only modest cross-datacenter bandwidth per machine. These results show that KVCache-efficient model architectures are necessary but not sufficient for cross-datacenter heterogeneous serving. What makes the deployment practical is the combination of model-side KVCache reduction with system-side selective offloading and bandwidth-aware scheduling. Together, they turn cross-datacenter PD disaggregation from an appealing idea into a realistic serving architecture.
\section{Background}
\label{sec:background}

\subsection{The Bandwidth Wall in Conventional PD Disaggregation}

Prefill-decode (PD) disaggregation has become the standard systems abstraction for large-scale LLM serving because it cleanly separates two fundamentally different phases of inference: prefill is dominated by arithmetic throughput, while decode is dominated by memory bandwidth. That separation improves utilization and enables phase-specific optimization. But it does not come for free. Once prefill and decode are placed on different nodes, every request must export its KVCache from the prefill side to the decode side, turning what was previously an on-device state transition into a cross-node transport problem. In practice, that transport requirement is what keeps today's PD deployments confined to a single data center and attached to RDMA-class scale-out networks.

\begin{table}[t]
\centering
\caption{Configurations of representative models. Type A denotes the linear-complexity block, and Type B denotes the quadratic-complexity full attention block.}
\label{tab:hybrid-config}
\begin{tabular}{lcccc}
\toprule
Model & Attention Type A & Attention Type B & A:B Ratio & Model Params \\
\midrule
Kimi Linear~\cite{team2025kimilinear} & KDA~\cite{team2025kimilinear} & MLA~\cite{liu2024deepseek} & 3:1 & 48B  \\
MiMo-V2-Flash~\cite{xiao2026mimo} & SWA~\cite{beltagy2020longformer} & GQA~\cite{ainslie2023gqa} & 5:1 & 309B \\
Qwen3.5-397B~\cite{qwen2026qwen3p5} & GDN~\cite{yang2025gated} & GQA & 3:1 & 397B \\
Ring-2.5-1T~\cite{inclusion2026ring25} & Lightning~\cite{qin2024lightning} & MLA & 7:1 & 1T \\
MiniMax-M2.5~\cite{minimax2026m25} & -- & GQA & -- & 229B \\
Qwen3-235B~\cite{yang2025qwen3} & -- & GQA & -- & 235B \\
\bottomrule
\end{tabular}
\end{table}

Under latency-sensitive serving constraints, KVCache produced by prefill instances can be shipped asynchronously to maximize compute utilization. To avoid GPU idling, the egress bandwidth $B_{\text{out}}$ of a prefill cluster must exceed the aggregate rate at which the cluster produces KVCache. Because this aggregate rate scales linearly with the number of instances, the binding constraint reduces to the \emph{KV throughput} of a single model instance, which we define as
\begin{equation}
\Phi_{\text{kv}}(l)=\frac{S_{\text{kv}}(l)}{T_{\text{prefill}}(l)},
\label{eq:kv-throughput}
\end{equation}
where $S_{\text{kv}}(l)$ is the KVCache size for a request of length $l$ and $T_{\text{prefill}}(l)$ is the corresponding prefill latency. The value of this metric depends largely on model architecture. Table~\ref{tab:hybrid-config} summarizes the configurations considered in this paper. For conventional dense-attention architectures, this transport demand is a dominant systems constraint. In standard Transformer-style attention, KVCache grows linearly with context length and can reach tens of gigabytes. Figure~\ref{fig:kv_throughput_minimax_m25} shows the KV throughput of MiniMax-M2.5, a representative dense model with GQA, at various input lengths. The bottleneck is stark: for a 32K-token request, a single MiniMax-M2.5 instance produces KVCache at roughly 60\,Gbps, requiring egress bandwidth that far exceeds the cross-datacenter Ethernet capacity of a typical machine. This is precisely why conventional PD disaggregation remains operationally tied to tightly integrated network domains. The network budget is so large that moving prefill and decode across looser interconnects, let alone across data centers, is simply not realistic.

That network coupling also prevents heterogeneous serving from scaling cleanly. Specialized chips already exist for each phase: hardware such as Rubin CPX targets prefill throughput, while LPU-style designs target decode bandwidth. Yet high-performance interconnects remain tightly coupled to machine form factors and deployment environments, so connecting unlike hardware at RDMA-class bandwidth generally requires bespoke engineering. Worse, once heterogeneous hardware is forced into a single tightly coupled cluster, the system inherits a fixed prefill-to-decode hardware ratio. In production traffic, request mix, request volume, and prefix-cache hit rate fluctuate continuously, so one side of the pipeline inevitably becomes overprovisioned while the other becomes the bottleneck. In a homogeneous cluster, any machine can be dynamically reassigned between prefill and decode roles as load shifts. A heterogeneous cluster offers no such flexibility: a chip specialized for prefill cannot serve decode and vice versa, leading to severe load imbalance and stranded capacity. The result is higher operational complexity and limited real-world adoption of heterogeneous PD beyond bespoke or low-throughput scenarios.

\begin{table}[t]
\centering
\begin{minipage}[c]{0.44\linewidth}
\centering
\includegraphics[width=\linewidth]{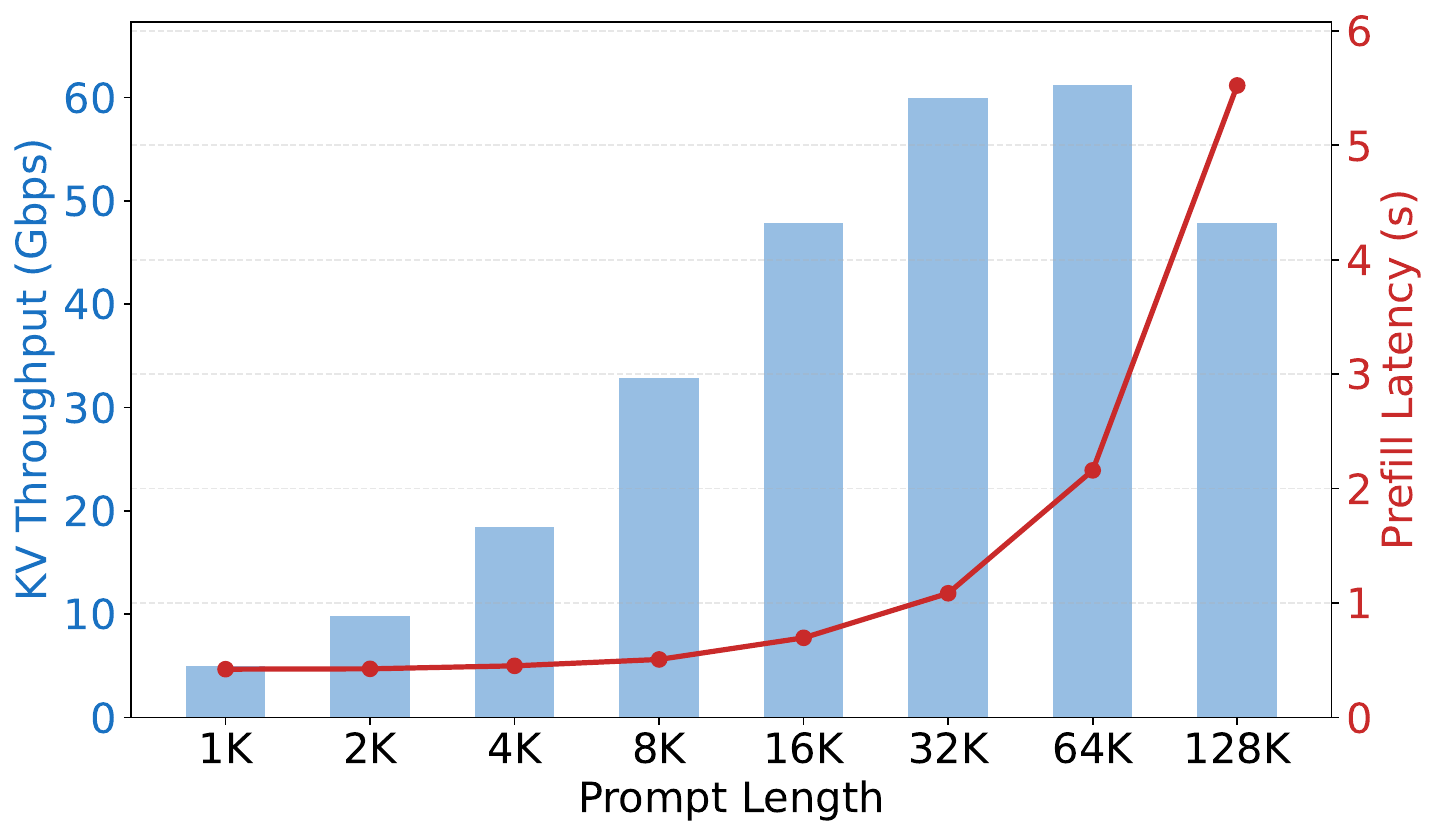}
\vspace{0cm}
\captionof{figure}{KV throughput of MiniMax-M2.5 on an 8$\times$H200 instance at various input lengths.}
\label{fig:kv_throughput_minimax_m25}
\end{minipage}
\hfill
\begin{minipage}[c]{0.52\linewidth}
\centering
\small
\vspace{0.4cm}
\begin{tabular}{lcc}
\toprule
Mechanism & Prefill Latency & KV Throughput \\
\midrule
GQA        & High & High \\
MLA        & High & Low  \\
Sparse Attention     & Low  & High \\
SWA       & Low  & Low  \\
Linear Attention & Low  & Low  \\
\bottomrule
\end{tabular}
\vspace{1cm}
\caption{Prefill latency and KV throughput characteristics of different attention mechanisms. Lower is better for both metrics.}
\label{tab:attention-mechanism}
\end{minipage}
\end{table}

\subsection{Hybrid Attention Changes the PD Deployment Boundary}
\label{sec:hybrid-attention-analysis}

\begin{table}[t]
\centering
\caption{KV throughput $\Phi_{\text{kv}}$ (Gbps) at various input lengths. All models are benchmarked on 8$\times$H200 with SGLang v0.5.9~\cite{sglang2025}.}
\label{tab:kv-throughput}
\footnotesize
\setlength{\tabcolsep}{4pt}
\begin{tabular}{l cc cccc}
\toprule
 & \multicolumn{4}{c}{Hybrid} & \multicolumn{2}{c}{Dense} \\
\cmidrule(lr){2-5} \cmidrule(lr){6-7}
Seq Len & Kimi Linear & MiMo-V2-Flash & Qwen3.5-397B & Ring-2.5-1T & MiniMax-M2.5 & Qwen3-235B \\
\midrule
1K   & 1.19 & 0.82 & 4.13 & 7.27 & 4.94 & 4.12 \\
8K   & 2.29 & 2.85 & 6.28 & 4.47 & 32.87 & 22.42 \\
32K  & 3.87 & 4.66 & 8.25 & 2.59 & 59.93 & 33.35 \\
128K & 4.88 & 4.71 & 7.47 & 1.46 & 47.82 & 21.50 \\
\bottomrule
\end{tabular}
\end{table}

What changes this picture is not a new scheduler alone, but a shift in model architecture. As LLMs move toward longer contexts, the cost of conventional MHA becomes increasingly untenable, prompting a broad transition toward KVCache-friendly designs. Table~\ref{tab:attention-mechanism} categorizes mainstream attention improvements along two dimensions: prefill latency ($T_{\text{prefill}}$) and KV throughput ($\Phi_{\text{kv}}$). Under long-context workloads, full attention mechanisms such as GQA and MLA retain quadratic complexity, resulting in high prefill cost. Sparse attention~\cite{child2019generating} reduces the amount of computation and can lower prefill latency, but it still requires transferring sequence-length-dependent KVCache to decode instances, leaving KV throughput as the dominant bottleneck. By contrast, linear attention and SWA maintain linear computation cost, while their bounded state size substantially reduces KV throughput.

A growing number of flagship open-source models adopt linear attention~\cite{team2025kimilinear,qwen2026qwen3p5,inclusion2026ring25,nvidia_nemotron_nano_v3_2025} or SWA~\cite{xiao2026mimo,huang2026step,agarwal2025gpt}, combining these mechanisms into hybrid stacks that interleave a small number of full-attention layers with a larger number of linear-complexity layers. Representative examples include Qwen3.5-397B at a 3:1 linear-to-full ratio, MiMo-V2-Flash at a 5:1 SWA-to-full ratio, and Ring-2.5-1T at a 7:1 linear-to-full ratio. In these architectures, only the full-attention layers produce KVCache that scales with sequence length, while the linear-complexity layers maintain fixed-size recurrent state whose footprint becomes negligible in the long-context regime.

Equation~\eqref{eq:kv-throughput} quantifies how model architecture governs the bandwidth demand of PD disaggregation. Table~\ref{tab:kv-throughput} benchmarks $\Phi_{\text{kv}}$ for several recent open-source dense and hybrid models. Compared with dense models of similar size, hybrid models exhibit a sharp reduction in $\Phi_{\text{kv}}$, meaning that each unit of compute generates far less state that must later traverse the network. At 32K tokens, MiMo-V2-Flash achieves a KV throughput of 4.66\,Gbps versus 59.93\,Gbps for MiniMax-M2.5, a 13$\times$ reduction. Qwen3.5-397B reaches 8.25\,Gbps versus 33.35\,Gbps for Qwen3-235B, a 4$\times$ reduction. The paper further notes that for Ring-2.5-1T, MLA contributes roughly a 4.5$\times$ compression over GQA, while the 7:1 hybrid ratio contributes another ${\sim}$8$\times$ reduction, yielding an overall KV memory saving of roughly 36$\times$.

The key systems implication is not merely lower inference cost, but a reduced KV throughput that shifts the deployable network boundary of PD disaggregation from RDMA-class fabrics to commodity Ethernet. In dense-attention models, prefill emits too much state too quickly, so the network becomes the hard coupling between phases. In hybrid models, prefill still performs substantial work, but the emitted KVCache is dramatically smaller. This does not make cross-datacenter PD free enough for indiscriminate transfer. Rather, it opens a qualitatively different operating regime in which cross-cluster KVCache transport becomes plausible for selected requests and therefore worth optimizing at the system level.

\subsection{From Intra-Datacenter PD to Prefill-as-a-Service Paradigm}
\label{sec:cluster-network-demand}

In intra-datacenter PD deployments, prefill and decode nodes communicate over high-bandwidth, fully meshed networks such as RDMA, so the network is far from being a bottleneck relative to prefill computation. In cross-cluster PD, however, the relationship between inter-cluster bandwidth and model KV throughput directly determines whether cross-datacenter KVCache is feasible. The cluster-level bandwidth requirement follows from the per-instance KV throughput. For an $N$-GPU prefill cluster, the minimum egress bandwidth is
\begin{equation}
B_{\text{out}}
  =\frac{N}{P}\cdot\frac{\mathbb{E}[S_{\text{kv}}]}{\mathbb{E}[T_{\text{prefill}}]}
  \approx\frac{N}{P}\cdot\Phi_{\text{kv}}(L_{\text{avg}}),
\label{eq:egress-bw}
\end{equation}
where $P$ is the parallelism degree (GPUs per instance) and $L_{\text{avg}}$ is the average \emph{uncached} input length of the requests actually offloaded to the \systemname cluster. Notably, $B_{\text{out}}$ depends not only on $\Phi_{\text{kv}}$, which is governed by model architecture and hardware, but also on $L_{\text{avg}}$, which is shaped by request length distribution, prefix-cache hit rate, and routing policy. This dependence is exactly why hybrid models alone are not enough. On the model side, adopting KVCache-friendly architectures reduces $\Phi_{\text{kv}}$. On the system side, selective offloading and a bandwidth-aware scheduler that accounts for bandwidth constraints and KVCache locality (\S\ref{sec:scheduling}) determine which requests consume the cross-datacenter budget in the first place, keeping $B_{\text{out}}$ within the available inter-datacenter bandwidth envelope.

To make the analysis concrete, we consider a prefill cluster comprising 512 H200 GPUs with $L_{\text{avg}}=32\text{K}$. Under this configuration, MiniMax-M2.5 and Qwen3 respectively require 3.8\,Tbps and 2.1\,Tbps of egress bandwidth, which effectively locks deployment into a tightly integrated single-cluster fabric. By contrast, models that employ hybrid architectures see their KV throughput drop by an order of magnitude, bringing the bandwidth demand to a level that modern inter-datacenter links can sustain. Ring-2.5-1T requires roughly 170\,Gbps of dedicated line capacity. Moreover, by routing even longer requests (e.g., 128K tokens) to the \systemname cluster for processing, the bandwidth demand falls further to below 100\,Gbps. Even at the scale of a 10,000-GPU datacenter, the aggregate egress bandwidth totals only about 1.8\,Tbps, well within the capacity of physical inter-datacenter links.

That is the systems turning point described before. Once KV throughput falls far enough, heterogeneous serving no longer has to be implemented solely as an awkward co-location of unlike accelerators behind the same RDMA island. Instead, prefill can be selectively externalized into standalone, compute-dense Prefill-as-a-Service clusters, while decode remains in conventional bandwidth-optimized PD clusters. The problem then shifts from ``how do we force heterogeneous hardware into one tightly coupled deployment?'' to ``how do we identify the requests worth offloading and transport their KVCache across loosely coupled clusters efficiently?'' In that sense, KVCache-friendly model architectures, together with selective offloading and bandwidth-aware scheduling, jointly enable heterogeneous PD disaggregation to scale beyond a single datacenter.
\section{Disaggregation over Cross-Datacenter KVCache}
\label{sec:design}

\subsection{Overview}

\begin{figure}[t]
\centering
\includegraphics[width=\linewidth]{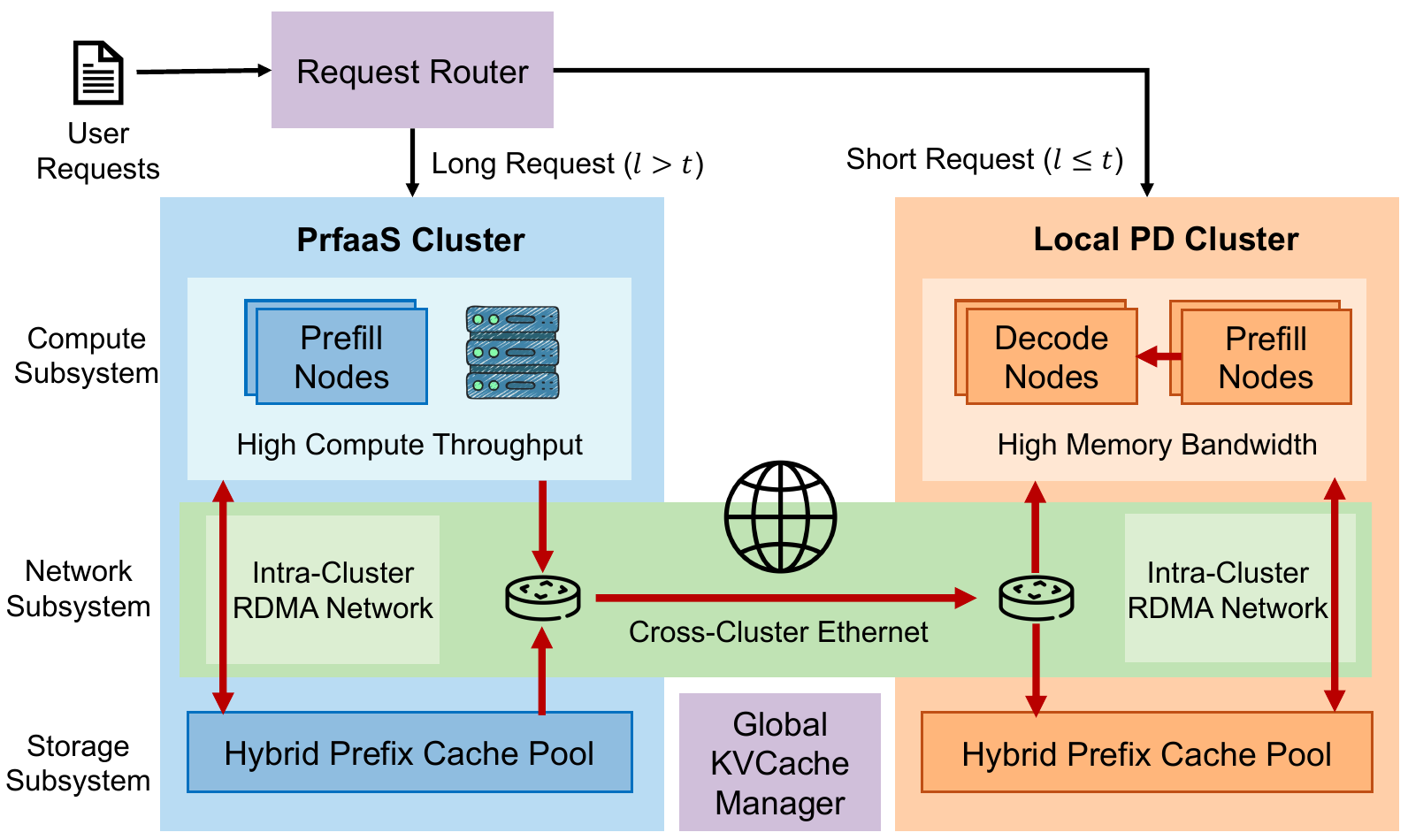}
\caption{Deployment topology of the \systemname-PD architecture.}
\label{fig:paas_pd_architecture}
\end{figure}

The core idea of cross-datacenter KVCache is not to externalize every prefill, but to selectively extend disaggregated LLM serving beyond the boundary of a single cluster when remote prefill acceleration is worth the transfer cost. We realize this vision through the \systemname-PD architecture, which leverages cross-datacenter KVCache to decouple prefill and decode across loosely connected clusters for requests whose long uncached prefills benefit most from faster compute. As shown in Figure~\ref{fig:paas_pd_architecture}, dedicated \systemname clusters perform compute-intensive long-context prefill on cost-effective, high-throughput accelerators and stream the resulting KVCache to local PD clusters via commodity Ethernet, while short or bandwidth-unfriendly requests remain on the local PD path.

The \systemname-PD architecture comprises three subsystems: compute, network, and storage. The compute subsystem consists of multiple clusters, each containing only homogeneous hardware, as different chip types are typically difficult to co-locate within the same facility. Clusters fall into two categories. \emph{Local PD clusters} perform PD-disaggregated serving and can complete inference for a request end to end. \emph{\systemname clusters} provide selective remote prefill capacity for requests whose incremental uncached length exceeds a routing threshold. After prefill, the resulting KVCache is transferred to a local PD cluster for decode. The network subsystem spans two tiers: intra-cluster networks use RDMA for latency-sensitive collective communication and PD KVCache transfers, while inter-cluster links rely on VPC peering or dedicated lines for cross-datacenter KVCache transfer. The storage subsystem resides within each cluster, building a distributed hybrid prefix cache pool (\S\ref{sec:hybrid-prefix-cache-pool}) across all machines. A global KVCache manager maintains KVCache metadata across all clusters. On top of these infrastructure components, a global scheduler routes requests to clusters and nodes based on request characteristics, network conditions, and cache distribution, maximizing end-to-end throughput under cross-cluster bandwidth constraints, as detailed in \S\ref{sec:scheduling}.

\subsection{Hybrid Prefix Cache Pool}
\label{sec:hybrid-prefix-cache-pool}

Prefix cache pools allow the serving system to offload KVCache to distributed host memory and SSDs within a cluster, substantially increasing the prefix cache hit rate. Conventional prefix cache pools, however, are designed for a single KVCache type and perform matching and eviction at the token or block level. In hybrid models, the recurrent states of linear attention or SWA layers are request-level: their size is independent of input length, and they can only be reused when the cached length matches exactly. In contrast, the KVCache of full attention layers are block-level: they grow linearly with input length and support partial prefix matching. This heterogeneity challenges the conventional all-layer-uniform KVCache storage paradigm.

\begin{figure}[t]
\centering
\includegraphics[width=0.8\linewidth]{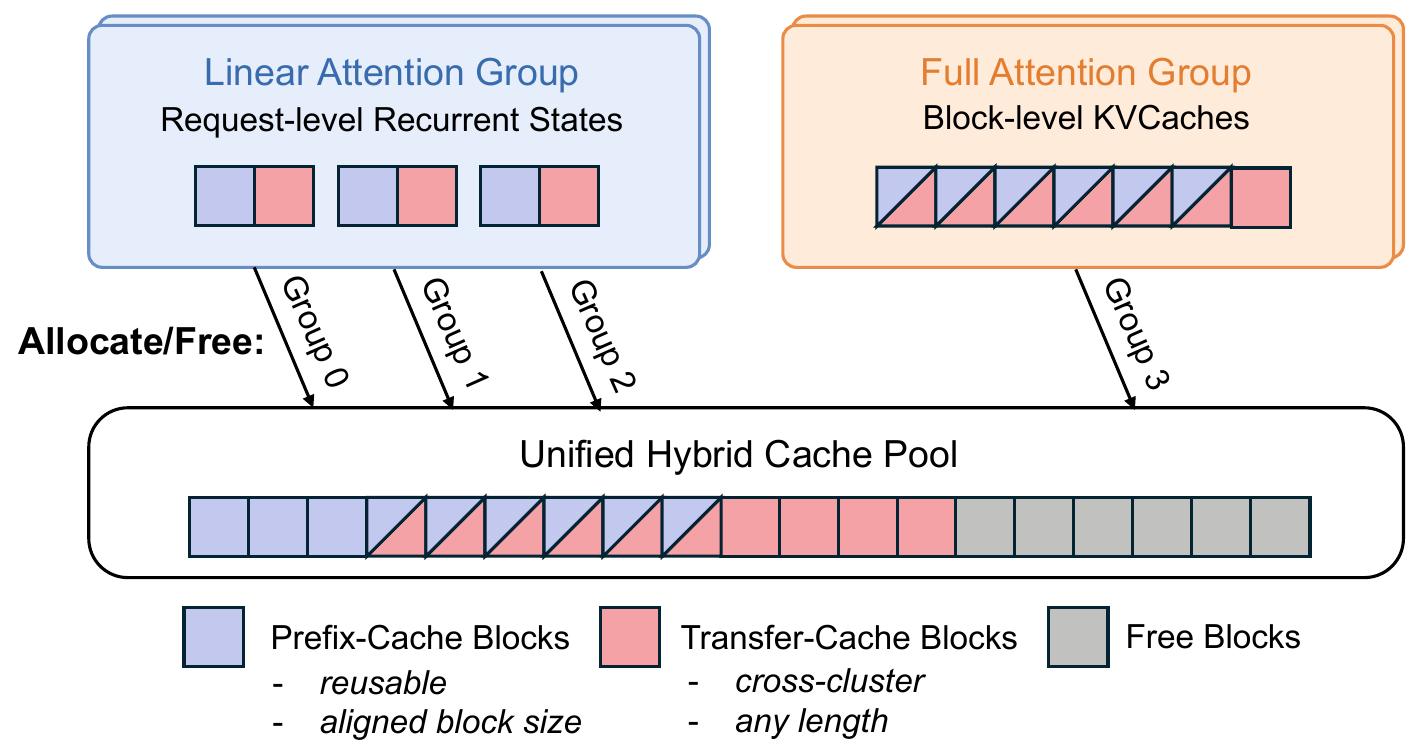}
\caption{Hybrid prefix cache pool. Linear states and full-attention KVCache are managed by separate groups backed by a unified block pool. Blocks are categorized as prefix-cache (intra-cluster only, block-aligned) or transfer-cache (cross-cluster, discarded after transfer).}
\label{fig:prefix_cache_pool}
\end{figure}

Based on vLLM's hybrid KVCache manager~\cite{vllm2025hybrid}, we build a hybrid prefix cache system tailored for cross-cluster KVCache transfer, as shown in Figure~\ref{fig:prefix_cache_pool}. Linear states and full-attention KVCache are managed by separate KVCache groups with aligned block sizes, allowing all groups to allocate and free blocks from a shared KVCache pool. On top of this shared pool, we partition cache blocks into two categories: prefix-cache blocks and transfer-cache blocks. Prefix-cache blocks must be fully populated before they can be reused across requests. Transfer-cache blocks hold the KVCache produced at the tail of a prefill request for PD-disaggregated transfer, and the cache pool discards them once the transfer completes.

When a new request arrives, the global KVCache manager computes prefix-match information for every cluster, and the request router uses this information to select the prefill cluster and the cache-affine node within it. Beyond routing, the KVCache manager also performs cache rebalancing to mitigate hotspots. When sufficient inter-cluster bandwidth is available, cross-cluster cache transfer is feasible as well, as discussed in \S\ref{sec:scheduling-strategy}.

\subsection{\systemname-PD Disaggregation}

Building on conventional intra-cluster PD disaggregation, we introduce \systemname clusters as a selective extension that increases deployment scalability and lowers cost without forcing all requests onto the cross-cluster path. A \systemname-PD system may contain several \systemname and local PD clusters, whose sizes and ratios are determined by hardware capabilities, network bandwidth, and request traffic.

Each \systemname cluster functions as a stateless KVCache producer whose effective throughput equals the minimum of its prefill computation rate and its network egress bandwidth. Not all requests benefit equally from being offloaded to a \systemname cluster. Short-context prefill is typically memory- or communication-bound rather than compute-bound, resulting in low arithmetic utilization that cannot fully exploit the compute-dense accelerators in \systemname clusters. We therefore retain prefill nodes within the local PD cluster and apply a length-based routing policy. Let $l$ denote the incremental prefill length of a request (excluding any cached prefix) and $t$ a routing threshold. When $l > t$, the request is routed to a \systemname cluster, and the resulting KVCache is transferred to a decode node upon completion. When $l \leq t$, the request is handled by a prefill node within the PD cluster. With the growing adoption of agentic workloads, the majority of requests are incremental prefills with prefix cache hits. For such requests, the global KVCache manager tracks the storage locations of all cached entries, ensuring that only the incremental portion is transferred across clusters. The scheduler must jointly consider cache affinity and cluster bandwidth when making routing decisions, as discussed in \S\ref{sec:scheduling-strategy}.

Realizing \systemname-PD disaggregation in practice requires stable, high-throughput Ethernet transport. Although hybrid model architectures substantially reduce the nominal bandwidth requirement, bursty traffic and uneven link utilization can still cause congestion, increasing queuing delay and degrading KVCache delivery efficiency. Our design therefore aims to smooth transfer traffic and sustain high link utilization without inducing persistent congestion. To this end, we combine layer-wise prefill pipelining to overlap KVCache generation with transmission, multi-connection TCP transport to fully utilize the available bandwidth, and congestion monitoring integrated with the scheduler to detect loss and retransmission signals early and prevent congestion accumulation.

\subsection{Modeling and Scheduling}
\label{sec:scheduling}

A \systemname-PD system comprises three roles: \systemname prefill, PD-P (the prefill nodes within a PD cluster), and PD-D (the decode nodes within a PD cluster). To derive scheduling policies, we construct an analytical throughput model that captures the compute and bandwidth constraints of each role and use it to guide both short-term routing and long-term resource allocation.

\subsubsection{Throughput Model}

\begin{table}[t]
\centering
\caption{Notation used in the \systemname-PD throughput model.}
\label{tab:notation}
\begin{tabular}{cl@{\hspace{1.5em}}cl}
\toprule
\multicolumn{2}{l}{Traffic} & \multicolumn{2}{l}{System} \\
\midrule
$\Lambda$ & Request arrival rate (throughput) & $N_{\text{\systemnamelowercase}}$ & \systemname prefill instances \\
$L$ & Uncached input length (r.v.) & $N_p$, $N_d$ & PD-P / PD-D instances \\
$t$ & Routing threshold & $B_{\text{out}}$ & \systemname egress bandwidth \\
$l_{\text{long}}$ & $\mathbb{E}[L \mid L > t]$, mean \systemname length & $\mathit{BS}_{\max}$ & Max decode batch size \\
$l_{\text{short}}$ & $\mathbb{E}[L \mid L \leq t]$, mean PD-P length & $T_{\text{prefill}}(l)$ & Prefill time for length $l$ \\
$p$ & $P(L > t)$, fraction to \systemname & $T_{\text{decode}}$ & Per-step decode time \\
$L_{\text{out}}$ & Mean output length & $\Theta_{\text{\systemnamelowercase}}$ & \systemname throughput (req/s) \\
$S_{\text{kv}}(l)$ & KVCache size for length $l$ & $\Theta_{\text{pd-p}}$, $\Theta_{\text{pd-d}}$   & PD-P / PD-D throughput (req/s) \\
\bottomrule
\end{tabular}
\end{table}

We model the steady-state throughput of the \systemname-PD system using the notation in Table~\ref{tab:notation}. For tractability, we approximate all \systemname requests by a representative length $l_{\text{long}} = \mathbb{E}[L \mid L > t]$ with service time $T_{\text{prefill}}(l_{\text{long}})$, and all PD-P requests by $l_{\text{short}} = \mathbb{E}[L \mid L \leq t]$ with service time $T_{\text{prefill}}(l_{\text{short}})$. Requests arrive uniformly at aggregate rate $\Lambda$. A fraction $p = P(L > t)$ of requests are routed to \systemname, and the remaining $1 - p$ are handled by PD-P.

Each \systemname request undergoes two pipelined phases: prefill computation and KVCache transfer. Through layer-wise prefill pipelining, the \systemname cluster throughput is determined by the slower of compute and egress transfer:
\begin{equation}
\label{eq:theta-paas}
\Theta_{\text{\systemnamelowercase}} = \min\!\left(\frac{N_{\text{\systemnamelowercase}}}{T_{\text{prefill}}(l_{\text{long}})},\; \frac{B_{\text{out}}}{S_{\text{kv}}(l_{\text{long}})}\right).
\end{equation}
Since intra-cluster RDMA bandwidth is not a bottleneck, PD-P throughput is determined solely by compute capacity:
\begin{equation}
\label{eq:theta-pdp}
\Theta_{\text{pd-p}} = \frac{N_p}{T_{\text{prefill}}(l_{\text{short}})}.
\end{equation}
The decode-stage (PD-D) throughput is
\begin{equation}
\label{eq:theta-pdd}
\Theta_{\text{pd-d}} = \frac{N_d \cdot \mathit{BS}_{\max}}{T_{\text{decode}} \cdot L_{\text{out}}},
\end{equation}
where $\mathit{BS}_{\max}$ and $T_{\text{decode}}$ are treated as SLO-governed constants.

\systemname and PD-P serve as upstream producers, each prefilling a disjoint subset of requests (fractions $p$ and $1{-}p$, respectively), while PD-D is the sole downstream consumer. Together, they form a converging pipeline:
\begin{center}
\begin{tikzpicture}[>=stealth, font=\small,
  box/.style={draw, rounded corners, inner sep=4pt, minimum height=1em}]
  \node[box] (req)  at (0,0)     {Request};
  \node[box] (paas) at (3, 0.8)  {\systemname Prefill};
  \node[box] (pdp)  at (3,-0.8)  {PD-P Prefill};
  \node[box] (pdd)  at (6.5,0)   {PD-D Decode};
  \draw[->] (req) -- node[above,sloped]{\footnotesize $p$} (paas);
  \draw[->] (req) -- node[below,sloped]{\footnotesize $1{-}p$} (pdp);
  \draw[->] (paas) -- node[above,sloped]{\footnotesize Ethernet} (pdd);
  \draw[->] (pdp)  -- node[below,sloped]{\footnotesize RDMA} (pdd);
\end{tikzpicture}
\end{center}
The end-to-end system throughput is limited by the slowest stage, accounting for the routing split:
\begin{equation}
\label{eq:lambda-max}
\Lambda_{\max} = \min\!\left(\frac{\Theta_{\text{\systemnamelowercase}}}{p},\; \frac{\Theta_{\text{pd-p}}}{1 - p},\; \Theta_{\text{pd-d}}\right).
\end{equation}

\subsubsection{Throughput-Optimal Configuration}

Given fixed hardware resources ($N_{\text{\systemnamelowercase}}$, $N_p{+}N_d$) and network bandwidth $B_{\text{out}}$, we seek two decision variables that maximize $\Lambda_{\max}$: the routing threshold $t$ (which determines $p$, $l_{\text{long}}$, and $l_{\text{short}}$) and the PD-cluster prefill-to-decode ratio $N_p / N_d$.

The threshold $t$ governs the trade-off between \systemname and PD-P load. Increasing $t$ restricts \systemname to longer requests, for which $T_{\text{prefill}}(l)$ grows faster than $S_{\text{kv}}(l)$ (near-quadratic prefill versus linear KVCache size). This lowers the per-instance KV throughput and eases bandwidth pressure. Conversely, decreasing $t$ floods \systemname with shorter requests whose high KV throughput is more likely to trigger the bandwidth bottleneck. The optimal $t$ balances \systemname and PD-P throughput so that both stages approach saturation simultaneously:
\begin{equation}
\label{eq:balance-paas-pdp}
\frac{\Theta_{\text{\systemnamelowercase}}}{p} = \frac{\Theta_{\text{pd-p}}}{1 - p}.
\end{equation}
For a fixed cluster size $N_p + N_d$ (the number of machines in a datacenter is constant in the short term), the ratio $N_p / N_d$ should balance the aggregate producer throughput against the consumer throughput. Too few prefill nodes starve the decode stage of KVCache, while too many leave prefill capacity idle. The optimal ratio satisfies
\begin{equation}
\label{eq:balance-pd}
\Theta_{\text{\systemnamelowercase}} + \Theta_{\text{pd-p}} = \Theta_{\text{pd-d}}.
\end{equation}
Equations~\eqref{eq:balance-paas-pdp} and~\eqref{eq:balance-pd} constrain two unknowns, $t$ and $N_p / N_d$. Because $\Theta_{\text{\systemnamelowercase}}/p$ decreases monotonically with $p$ (and hence with decreasing $t$) while $\Theta_{\text{pd-p}}/(1{-}p)$ increases, a grid search over $t$ and $N_p / N_d$ efficiently finds the optimal operating point.

\subsubsection{Dual-Timescale Scheduling}
\label{sec:scheduling-strategy}

The scheduling is central to making cross-datacenter disaggregation practical rather than merely architecturally possible. In theory, hybrid model architectures reduce KV throughput enough that commodity Ethernet links can sustain cross-cluster transfers, and the steady-state analysis above can maximize cluster throughput by optimizing $t$ and $N_p / N_d$. In practice, however, traffic variations and burstiness can cause transient congestion and queue buildup at the \systemname egress. Moreover, clusters maintain large prefix caches whose bulk transfers can further strain cross-cluster links. To handle this dynamic environment, we design a dual-timescale scheduling algorithm that treats cross-cluster bandwidth and throughput as the primary constraints, separating the factors governing request routing and resource allocation into short-term and long-term categories with a dedicated strategy for each.

\paragraph{Short-term: bandwidth- and cache-aware routing.}
The \systemname cluster has a bandwidth-imposed throughput ceiling $B_{\text{out}} / S_{\text{kv}}(l_{\text{long}})$. As the cluster approaches this ceiling, congestion builds at the egress link. The scheduler therefore continuously monitors the \systemname egress utilization and request queue depth. When utilization approaches a threshold or queuing surges, a short-term routing adjustment is triggered.

At initialization or upon a policy update, the scheduler profiles current compute capacity and egress bandwidth, then searches for the optimal threshold $t$ based on the incremental prefill-length distribution after prefix matching. By routing only sufficiently long requests to \systemname, the scheduler reduces per-request bandwidth demand and avoids congestion near the bandwidth ceiling.

For requests with prefix cache hits, the scheduler must jointly consider cache affinity and bandwidth availability. Let $l_{\text{total}}$ denote the total input length of a request, and let $l_{\text{\systemnamelowercase}}$ and $l_{\text{pd}}$ denote the cached prefix lengths in the \systemname and PD clusters, respectively. The routing depends on whether bandwidth or compute is the binding constraint.
When bandwidth is scarce, prefix caches in each cluster are evaluated independently: if $l_{\text{total}} - l_{\text{pd}} \leq t$, the request is prefilled locally in PD-P, and otherwise offloaded to \systemname. 
When bandwidth is abundant, compute becomes the scarce resource, and cross-cluster cache transfer can reduce redundant computation. The scheduler considers the best cache across all clusters by letting $l_{\text{prefix}} = \max(l_{\text{\systemnamelowercase}},\, l_{\text{pd}})$; if $l_{\text{total}} - l_{\text{prefix}} \leq t$, the request is prefilled in PD-P; otherwise it goes to \systemname. When the cluster with the longer cache differs from the compute cluster, a cross-cluster cache transfer is performed.

\paragraph{Long-term: traffic-driven allocation re-optimization.}
Over longer timescales, shifts in traffic volume create persistent imbalances between pipeline stages. 
When $\Theta_{\text{\systemnamelowercase}} + \Theta_{\text{pd-p}} \ll \Theta_{\text{pd-d}}$, prefill is the system bottleneck, and when $\Theta_{\text{\systemnamelowercase}} + \Theta_{\text{pd-p}} \gg \Theta_{\text{pd-d}}$, decode is. The scheduler identifies the binding constraint by monitoring queue depth and utilization at each stage. Since traffic variations at this timescale are gradual and often periodic, the scheduler periodically re-evaluates load balance and converts nodes between prefill and decode roles within the PD cluster, adjusting $N_p$ and $N_d$ to restore the optimality conditions of Equations~\eqref{eq:balance-paas-pdp} and~\eqref{eq:balance-pd}. As instance counts change, the routing threshold $t$ is re-optimized accordingly.

\section{Case Study: Bandwidth Demand of \systemname-PD Architecture}
\label{sec:case_study}

In this section, we use an internal 1T-parameter hybrid-architecture model as a case study to evaluate whether selective \systemname offloading can keep cross-datacenter KVCache transfer within a realistic bandwidth budget while improving system throughput under realistic hardware and deployment settings. Following the notation in Table~\ref{tab:notation}, we apply the profiling-based throughput model from \S\ref{sec:scheduling} to solve for two key parameters that maximize system throughput: the routing threshold $t$ and the prefill-to-decode instance ratio $N_p / N_d$ within the local PD cluster. Under the resulting configuration, the heterogeneous \systemname-PD deployment achieves 54\% higher throughput and 64\% lower P90 TTFT than a homogeneous PD-only baseline, and 32\% higher throughput than a naive heterogeneous deployment without scheduling. The average \systemname cluster egress is only 13\,Gbps, well within the Ethernet capacity.

\subsection{Setup}

\begin{table}[t]
\centering
\caption{KVCache size $S_{\text{kv}}$, prefill latency $T_{\text{prefill}}$, and KV throughput $\Phi_{\text{kv}}$ of the internal 1T hybrid model at various input lengths. Prefill latency is benchmarked on 8$\times$H200 with in-house vLLM~\cite{vllm2025}.}
\label{tab:kv-throughput-case-study}
\setlength{\tabcolsep}{20pt}
\begin{tabular}{lccc}
\toprule
Seq Len & KVCache Size & Prefill Latency & KV Throughput \\
\midrule
1K   & 190.8\,MiB  & 0.44\,s  & 3.61\,Gbps \\
8K   & 308.9\,MiB  & 0.72\,s  & 3.59\,Gbps \\
32K  & 701.3\,MiB  & 1.84\,s  & 3.19\,Gbps \\
128K & 2316.3\,MiB & 7.40\,s  & 2.62\,Gbps \\
\bottomrule
\end{tabular}
\end{table}

We deploy two clusters connected by a VPC network, providing an aggregate cross-cluster bandwidth of approximately 100\,Gbps. The \systemname cluster consists of 32 H200 GPUs with higher compute throughput, dedicated to long-context prefill requests with $L > t$. The local PD cluster consists of 64 H20 GPUs operating in conventional PD-disaggregated mode with 800\,Gbps RDMA interconnect per node, where the prefill-to-decode ratio can be adjusted according to request traffic. Note that although H200 and H20 have different price points, the \systemname cluster only requires high compute throughput. In production deployments, cost-effective accelerators with comparable compute capability can serve as substitutes. As a baseline, we use a homogeneous PD cluster of 96 H20 GPUs.

To reflect realistic serving requirements, the workload uses an internal hybrid model with 1T parameters whose architecture follows Kimi Linear~\cite{team2025kimilinear}, employing an interleaved KDA:MLA layer structure at a 3:1 ratio. The model is deployed at 8 GPUs per instance and profiled separately for prefill and decode using in-house vLLM. Table~\ref{tab:kv-throughput-case-study} lists the KVCache size $S_{\text{kv}}$, prefill latency $T_{\text{prefill}}$, and KV throughput $\Phi_{\text{kv}}$ for this model at various input lengths.

Request input lengths follow a truncated log-normal distribution ($\mu = 9.90$, $\sigma = 1.00$, truncated to $[128,\, 128\text{K}]$) with a mean of approximately 27K tokens, reflecting the long-context distribution characteristic of real-world workloads. The output length is fixed at 1024 tokens, and the SLO (excluding speculative decoding) is set to 40 tokens/s. All throughput and bandwidth results reported below are derived by feeding the measured profiling data into the throughput model of \S\ref{sec:scheduling}.

\subsection{Throughput Modeling and Solution}

\begin{figure}[t]
\centering
\begin{subfigure}[t]{0.48\linewidth}
    \centering
    \includegraphics[width=\linewidth]{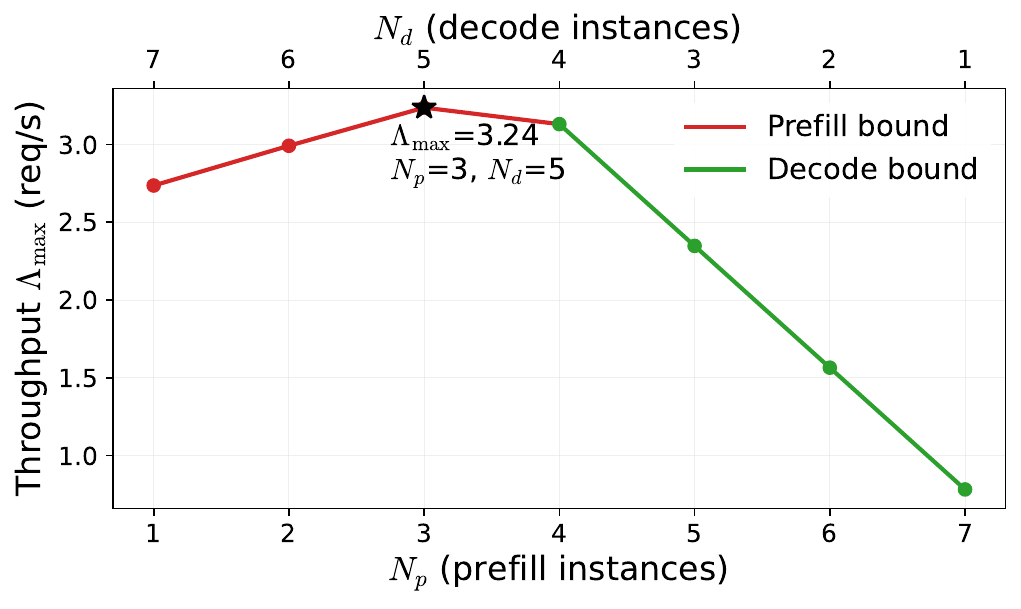}
    \caption{Search over prefill/decode allocation.}
    \label{fig:allocation-sensitivity}
\end{subfigure}
\hfill
\begin{subfigure}[t]{0.48\linewidth}
    \centering
    \includegraphics[width=\linewidth]{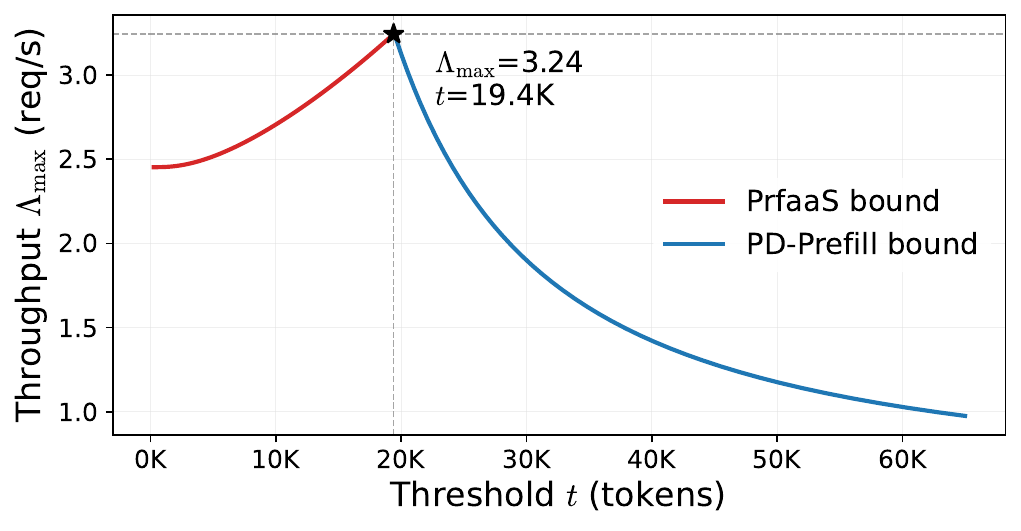}
    \caption{Search over routing threshold $t$.}
    \label{fig:threshold-sensitivity}
\end{subfigure}
\caption{Illustration of the grid search process for the two optimization variables. (a) fixes $t$ at the optimum and searches over the prefill/decode instance split within the local PD cluster. (b) fixes $N_p = 3$, $N_d = 5$ and searches over $t$.}
\label{fig:sensitivity}
\end{figure}

Using the \systemname-PD throughput model from \S\ref{sec:scheduling}, we optimize the routing threshold $t$ and the PD-cluster prefill/decode allocation to maximize $\Lambda_{\max}$. We solve them by exhaustive two-dimensional grid search. The optimal configuration is listed in the second column of Table~\ref{tab:paas-pd-comparison}.

Figure~\ref{fig:sensitivity} illustrates the search process, fixing one variable and searching over the other. Figure~\ref{fig:allocation-sensitivity} shows that, for a fixed threshold $t$ (and hence fixed $\Theta_{\text{\systemnamelowercase}}$), the system throughput peaks when prefill and decode throughput are approximately balanced. The optimal allocation in the local PD cluster is $N_p = 3$ and $N_d = 5$. Figure~\ref{fig:threshold-sensitivity} fixes $N_p = 3$ and $N_d = 5$, therefore the overall $\Lambda_{\max}$ is determined by $\min(\Theta_{\text{\systemnamelowercase}} / p,\; \Theta_{\text{pd-p}} / (1{-}p))$. The optimum occurs at the intersection of the two curves, yielding $t = 19.4$K. At this operating point, approximately 50\% of all requests (the longer ones) are offloaded to the \systemname cluster, fully utilizing the high-compute-throughput accelerators.

\begin{table}[t]
\centering
\caption{Comparison of optimal configurations across \systemname-PD, homogeneous PD, and naive heterogeneous PD deployments.}
\label{tab:paas-pd-comparison}
\begin{tabular}{lccc}
\toprule
Metric & \systemname-PD & Homogeneous PD & Naive Heterogeneous PD \\
\midrule
Threshold $t$ & 19.4K & --- & --- \\
$N_{\text{\systemnamelowercase}}$ / $N_p$ / $N_d$ & 4 / 3 / 5 & --- / 9 / 3 & 4 / --- / 8 \\
Mean / P90 TTFT (s) & 2.22 / 3.51 & 4.44 / 9.73 & 1.74 / 3.51 \\
$\Theta_{\text{\systemnamelowercase}}$ / $\Theta_{\text{pd-p}}$ / $\Theta_{\text{pd-d}}$ (req/s) & 1.61 / 1.64 / 3.91 & --- / 2.11 / 2.35 & 2.45 / --- / 6.25 \\
$\Lambda_{\max}$ (req/s) & 3.24 & 2.11 & 2.45 \\
Ratio & \textbf{1.54$\times$} & 1.00$\times$ & 1.16$\times$ \\
\bottomrule
\end{tabular}
\end{table}

\subsection{Result Analysis}

\subsubsection{Cross-Datacenter Bandwidth Utilization}

A key prerequisite for the \systemname-PD architecture is that the inter-cluster network link can sustain the KVCache transfer demand. We evaluate this by measuring the KV throughput of the \systemname cluster under the deployed workload distribution.

With the routing threshold set to $t = 19.4$K, 49.6\% of requests are routed to \systemname, and the offloaded subset has $\mathbb{E}[L \mid L > t] \approx 44$K tokens. The aggregate \systemname egress load is therefore approximately 13\,Gbps, consuming only 13\% of the Ethernet link. This confirms that the KVCache of a hybrid-architecture model can be transported over commodity Ethernet with substantial headroom with proper scheduling. By contrast, conventional Transformer models with full KV heads would produce significantly larger KVCache volumes, potentially requiring RDMA-class interconnects.

\subsubsection{Comparison with Homogeneous PD}

We apply the same throughput modeling methodology to the homogeneous PD baseline of 96 H20 GPUs. The results are shown in the third column of Table~\ref{tab:paas-pd-comparison}. In the homogeneous cluster, the system throughput is also maximized when prefill and decode throughput are balanced, yielding an optimal allocation of 9 prefill instances and 3 decode instances.

Thanks to the superior compute throughput of the \systemname cluster, the \systemname-PD configuration requires two fewer prefill instances in the local PD cluster, freeing capacity for additional decode slots. The overall system throughput improves by 54\%.

Another key benefit of \systemname-PD is the reduction in TTFT, particularly for long-context requests. In the homogeneous baseline, long requests compete with short requests for prefill capacity, inflating queuing delays. In \systemname-PD, long requests are offloaded to the dedicated high-throughput \systemname cluster, where prefill completes significantly faster than in the PD cluster even after accounting for cross-cluster transfer latency. As shown in Table~\ref{tab:paas-pd-comparison}, the mean and P90 TTFT are reduced by 50\% and 64\%, respectively, compared to the homogeneous baseline.

\subsubsection{Comparison with Naive Heterogeneous PD}

The comparison with homogeneous PD demonstrates the fundamental performance advantage of heterogeneous deployment. The comparison with naive heterogeneous PD further highlights the importance of scheduling in heterogeneous systems. In the naive heterogeneous PD configuration, no scheduling optimization is applied: all prefill is assigned to H200 GPUs and all decode to H20 GPUs, without length-based routing or load balancing. As shown in Table~\ref{tab:paas-pd-comparison}, the naive heterogeneous PD achieves only $1.16\times$ throughput over the homogeneous baseline, a 25\% reduction compared to \systemname-PD. This degradation stems from the severe imbalance between prefill and decode throughput, and more fundamentally from treating heterogeneous prefill as a universal path rather than selectively offloading only the requests that benefit most from \systemname.

\subsection{Summary}

This case study demonstrates that for hybrid-architecture models, cross-datacenter KVCache becomes practical when it is paired with selective \systemname offloading rather than applied indiscriminately.

Regarding feasibility, the KVCache transfer of a hybrid model consumes only 13\% of a 100\,Gbps Ethernet link. This is well within reach of commodity Ethernet infrastructure and far below the bandwidth demands of RDMA interconnects, confirming the viability of cross-datacenter KVCache transfer.

Regarding effectiveness, the \systemname-PD configuration (32 H200 GPUs for \systemname, 64 H20 GPUs for local PD) achieves 54\% higher throughput and 64\% lower P90 TTFT over a 96-H20 homogeneous PD-only baseline; at equal cost, the throughput gain is approximately 15\%. These gains stem from offloading compute-intensive long-context prefill to high-throughput \systemname accelerators. We note that H200 and H20 serve here as a representative hardware pair, not the sole viable combination. Cost-effective prefill-specialized chips can further reduce deployment cost in production.

Besides, the \systemname cluster is currently compute-bound with ample bandwidth headroom. Under larger-scale deployments or higher-bandwidth dedicated lines, the \systemname cluster can be further expanded to yield additional throughput gains. For example, in IDC-scale deployments with thousands of \systemname GPUs, the aggregate egress bandwidth required for KVCache transfer remains on the order of 1\,Tbps, well within the capacity of modern datacenter fabrics, enabling further improvements in throughput and resource efficiency.
\section{Discussion}
\label{sec:discussion}

Cross-datacenter KVCache extends PD disaggregation from a single tightly coupled cluster to loosely connected heterogeneous clusters. Its practicality depends on coordinated progress across model architecture, system design, and hardware. In this section, we discuss how these trends reinforce one another and what they imply for next-generation LLM serving systems.

\paragraph{KVCache-friendly model architecture.}
As context windows continue to grow, KVCache increasingly dominates inference cost in storage and transfer. Architectural techniques such as MLA, sliding window attention, and linear attention have already shown that KVCache size can be reduced substantially without sacrificing model capability. Going forward, model co-design is likely to optimize not only FLOPs but also KVCache transfer volume. These improvements directly reduce the latency and bandwidth cost of cross-datacenter KVCache, expanding the range of deployments in which \systemname-PD is cost-effective.

\paragraph{KVCache compression and reuse.}
Beyond architectural innovations, a growing body of work reduces LLM inference cost through KVCache compression and reuse at the algorithm or system level. Methods such as H2O~\cite{zhang2023h2o} and KIVI~\cite{liu2024kivi} selectively evict or quantize KVCache entries to shrink memory footprints. CacheGen~\cite{liu2024cachegen} applies conventional compression techniques to reduce KVCache transfer volume, while CacheBlend~\cite{yao2025cacheblend} and FusionRAG~\cite{wang2026prefix} enable reuse of approximately matched KVCache across requests. Together, these techniques complement KVCache-friendly model design by reducing effective memory pressure and network traffic, thereby making cross-datacenter KVCache more robust under production workloads.

\paragraph{Phase-specialized inference hardware.}
The disaggregation between prefill and decode is also becoming visible in hardware design. Prefill is compute-intensive, whereas decode is dominated by memory bandwidth. Recent chip roadmaps are increasingly phase-specialized: NVIDIA Rubin CPX~\cite{nvidia2025rubincpx} emphasizes compute throughput for prefill, while chips such as the LPU~\cite{groq2025lpu} and Taalas HC1~\cite{taalas2025hc1} feature extremely high memory bandwidth for fast decode. Cross-datacenter KVCache fits this trend naturally. It removes the requirement that heterogeneous chips be deployed within the same tightly coupled network domain, allowing operators to size prefill and decode clusters independently and to deploy each phase on the hardware best suited to it.
\section{Related Work}
\label{sec:related_work}

System optimization for online LLM inference has gradually shifted from monolithic single-cluster engines toward disaggregated, heterogeneity-aware, and KVCache-centric pipelines. On one hand, disaggregated serving splits each request into a compute-intensive prefill phase and a memory-bandwidth-intensive decode phase to reduce inter-phase interference and allow independent scaling. Splitwise~\cite{patel2024splitwise} and DistServe~\cite{zhong2024distserve} formulate PD disaggregation from cost/power and goodput perspectives, respectively, with corresponding deployment, scheduling, and placement strategies, showing that PD disaggregation can simultaneously improve throughput and reduce cost under appropriate SLO and hardware constraints. On the other hand, as cluster hardware and interconnects become increasingly heterogeneous, Helix~\cite{mei2025helix}, Hetis~\cite{mo2025hetis}, and LLM-PQ~\cite{zhao2024llmpq} incorporate heterogeneous GPUs/networks and phase-level differences into the optimization space to achieve throughput or latency gains through phase-specialized hardware placement. Meanwhile, DynamoLLM~\cite{stojkovic2025dynamollm} and FREESH~\cite{he2025freesh} emphasize system-level resource reconfiguration and cross-domain scheduling from the perspective of energy, cost, and carbon efficiency while meeting serving SLOs. More critically, KVCache has evolved from per-request ephemeral state into a first-class system resource: Mooncake~\cite{qin2024mooncake} introduces a global KVCache pool to improve cross-node and cross-request reuse; CacheBlend~\cite{yao2025cacheblend} and FusionRAG~\cite{wang2026prefix} significantly reduce TTFT through non-prefix KVCache fusion reuse; KIVI~\cite{liu2024kivi}, KVQuant~\cite{hooper2024kvquant}, and H2O~\cite{zhang2023h2o} further shrink KVCache volume via quantization or importance-based eviction to improve long-context serviceability. However, no prior work jointly optimizes cross-datacenter prefill offloading, heterogeneous deployment, and bandwidth/cache-aware scheduling within a unified system. This paper approaches the problem from a cross-datacenter KVCache perspective, combining disaggregated inference with heterogeneous resource orchestration to build a low-cost, high-throughput serving system.

\section{Conclusion}
\label{sec:conclusion}

To address the practical deployment challenges of heterogeneous disaggregated inference, we propose the concept of cross-datacenter KVCache, extending disaggregated serving from single homogeneous clusters to cross-cluster heterogeneous deployments. On this basis, we design the \systemname-PD disaggregation architecture, which augments system serving throughput at low cost through heterogeneous \systemname clusters connected via commodity Ethernet. We envision that the cross-datacenter KVCache paradigm will co-evolve with next-generation models, hardware, and networks to enable highly efficient LLM serving at scale.

\bibliographystyle{plain}
\bibliography{references}

\end{document}